\documentclass[fleqn,10pt]{wlscirep}
\usepackage[utf8]{inputenc}
\usepackage[T1]{fontenc}
\usepackage{bm}
\usepackage{eucal}
\usepackage{amsmath}
\usepackage[]{graphicx}
\usepackage{subfig}

\usepackage{float}
\usepackage{multirow}

\newcommand{\scn}{SchNet}
\newcommand{\dmn}{DimeNet}
\DeclareMathOperator{\sigmoid}{sigmoid}
\newcommand*\samethanks[1][\value{footnote}]{\footnotemark[#1]}

\title{Atomistic Graph Neural Networks for metals: Application to bcc iron}
\author[1]{Lorenzo Cian\thanks{These authors contributed equally: L. Cian and G. Lancioni.}}
\author[1]{Giuseppe Lancioni\samethanks[1]}
\author[2]{Lei Zhang}
\author[1]{Mirco Ianese}
\author[1]{Nicolas Novelli}
\author[1,3]{Giuseppe Serra}
\author[2,3]{Francesco Maresca}
\affil[1]{Artificial Intelligence LABoratory (AILAB), University of Udine, Udine, Italy}
\affil[2]{ENgineering and TEchnology institute Groningen (ENTEG), University of Groningen, Groningen, The Netherlands}

\affil[3]{Correspondence to: giuseppe.serra@uniud.it and f.maresca@rug.nl}

\begin{abstract}

The prediction of the atomistic structure and properties of crystals including defects based on {\it ab-initio} accurate simulations is essential for unraveling the nano-scale mechanisms that control the micromechanical and macroscopic behaviour of metals. Density functional theory (DFT) can enable the quantum-accurate prediction of some of these properties, however at high computational costs and thus limited to systems of $\sim 1,000$ atoms. In order to predict with quantum-accuracy the mechanical behaviour of nanoscale structures involving from thousands to several millions of atoms, machine learning interatomic potentials have been recently developed. Here, we explore the performance of a new class of interatomic potentials based on Graph Neural Networks (GNNs), a recent field of research in Deep Learning. Two state-of-the-art GNN models are considered, {\scn} and {\dmn}, and trained on an extensive DFT database of ferromagnetic bcc iron. We find that the {\dmn} GNN Fe potential including three-body terms can reproduce with DFT accuracy the equation of state and the Bain path, as well as defected configurations (vacancy and surfaces). To the best of our knowledge, this is the first demonstration of the capability of GNN of reproducing the energetics of defects in bcc iron. We provide an open-source implementation of {\dmn} that can be used to train other metallic systems for further exploration of the GNN capabilities.

\end{abstract}
\begin{document}

\flushbottom
\maketitle

\thispagestyle{empty}






\section*{Introduction}
Recently, Graph Neural Networks (GNNs) have become one of the most active research fields in Artificial Intelligence \cite{ZHOU202057}. GNNs are a class of Deep Learning methods introduced to analyze data which display a graph structure. Graphs represent the topology of a great variety of data structures in which objects (nodes) are connected with each other by some kind of relation (edges). 
Due to the very general nature of graphs, applications of GNNs are found in very different contexts, such as computer vision\cite{8578815, 6128754,DBLP:journals/corr/abs-1711-11575,9010993}, natural language processing\cite{Yao_Mao_Luo_2019, huang-etal-2019-text,bastings-etal-2017-graph, marcheggiani-etal-2018-exploiting},  social sciences \cite{9139346,10.3389/fdata.2019.00002}, and natural sciences including biology \cite{NIPS2017_f5077839}, particle physics \cite{Shlomi_2021} and astrophysics \cite{inproceedings}. 

The topology of a graph can also reflect that of atomistic crystal structures: indeed, a graph can be generated by connecting each atom (nodes) with its neighbors (edges), within a specified cutoff radius \cite{doi:10.1063/1.5126336}. The Message Passing Neural Networks framework (MPNN) \cite{pmlr-v70-gilmer17a} has been introduced as a common GNN paradigm for atomistic structures in quantum chemistry applications. 

Within an atomistic GNN, the atoms and their connections are associated with numerical lists of ``features'', also named {\it embedding} vectors.
Features are updated by the Message Passing framework, which is a two-step process. In the first step, each atom receives a message that is an aggregate of its neighbour's embeddings. In the second step, an updated embedding of the atom is evaluated, by means of a function that depends on the message and on the current atomic embedding. 
By iterating this scheme $n$ times, each atom will receive messages from atoms that are distant up to $n$ connections, thus accounting for long-range interactions.

A GNN model for atomistic graphs is therefore determined: (i) by the nature and the size of the embeddings, which convey the informative content of the specific atomistic system and (ii) by the operations it executes on them, i.e. the procedure used to aggregate and update embeddings.
Once the aforementioned characteristics are defined, the model can be trained to predict the system potential energy surface (PES).

A number of GNN models have been proposed in the recent years to model atomistic systems. Most of them were first introduced in molecular research and further applied to crystals. 
Deep Tensor Neural Networks (DTNN) \cite{Sch_tt_2017} and PhysNet \cite{Unke_2019} aggregate the atomic embeddings by means of {\it filters} that ensure that the resulting message changes smoothly with respect to small changes of the interatomic distances. The main difference between DTNN and PhysNet lies in how distances are represented and how messages are aggregated. 
Crystal Graph Convolutional Neural Networks (CGCNN) \cite{Xie_2018} were explicitly developed to deal with materials displaying a crystal structure, such as metals.  
Unlike DTNN and PhysNet, CGCNN considers both atomic (node) and edge embeddings; however, distances are not regularized with continuous functions: the range of the distances is partitioned in ten equally spaced segments, and interatomic distances are encoded within a single vector in which all components are zero but the one associated with the matching segment. Thus, this model lacks the ability to smoothly change the embeddings with respect to small displacements of the atomic positions. 
\scn \cite{10.5555/3294771.3294866} is based on DTNN and introduces continuous filter convolutions: distances are used as input of a neural layer that generates a continuous mapping to an embedding space. In an updated version \cite{doi:10.1063/1.5019779}, periodic boundary conditions (PBCs) have been introduced, and the model has been applied to the prediction of formation energies of bulk crystals. Also inspired by {\scn} and sharing its overall architecture, the ``Neural message passing with edge updates'' \cite{jorgensen2018neural} uses both node and edge embeddings in the form of a concatenation of the two connected atoms embeddings. 
This makes edge embeddings directional as they depend on the order of the concatenated elements.
MatErials Graph Network (MEGNet) \cite{Chen_2019} leverages a similar scheme, by incorporating both directional edge and node updates, while also introducing a global state vector which stores the molecule/crystal-level or state attributes, e.g. the temperature of the system. Updates of atoms, bonds and global state vector are performed in a sequence.
All these approaches employ filters that rely only on the distance between pairs of atoms to aggregate and update the atomic embeddings.

It is well-established that classical, empirical interatomic potentials \cite{PhysRevB.29.6443} that rely on pairwise interactions often fail to reproduce structural changes \cite{Lee2012AtomisticMO} and some crucial properties of dislocations in metals \cite{2018npjCM...4...69M}. In the case of phase transitions, the addition of directionality, i.e. angular dependence of the interatomic potential, as well as second nearest-neighbor interactions, has lead to the improved qualitative reproduction of quantum-mechanical PES \cite{Lee2012AtomisticMO}.

Within the context of GNNs, there is a remarkable shortage of approaches that rely also on the angle between edges connecting atomic pairs. Embeddings of edges connecting triplets of atoms convey the angular information, and once they are updated via the message passing scheme, they can be used to update the atomic embeddings. 
With this aim, {\dmn}\cite{Klicpera2020Directional} also leverages the Directional Message (hence the name) by considering the direction of the pairwise connections and by introducing the angle between two edges connected within atomic triplets. 
{\dmn} employs a continuous filter convolution by expanding both distances and angles in a Bessel-Fourier basis. However, to date, {\dmn} has been applied merely to isolated molecules and has not been investigated to model crystals such as metals.

Although GNNs have been scarcely explored in the context of interatomic potentials for metals, they introduce a number of advantages with respect to other ML methods \cite{doi:10.1063/1.5126336}. First, interactions among neighbouring atoms are straightforwardly modeled as pair-wise connections. Previous ML approaches need to introduce specific geometrical descriptors of the environment around atoms (within a cut-off radius), such as atom-centered symmetry functions in the Neural Networks Potentials \cite{PhysRevLett.98.146401}, or bispectrum components and then smooth overlap of atomic positions (SOAP) \cite{PhysRevB.87.184115} in Gaussian Approximation Potentials (GAP) \cite{PhysRevLett.104.136403}. Second, iterating the process makes the model able to consider the contributions of distant atoms, so as to mimic the influence of long-range interactions beyond the cut-off distance that limits pairwise interactions. This can be easily achieved by stacking message-passing layers in the network. Previous ML approaches either lack these long-range contributions or account for them by adding extra long-range terms to the total energy, e.g. for electrostatic interactions \cite{https://doi.org/10.1002/qua.24890}. 
Third, the GNN approach guarantees scalability of the system, as the pair-wise nature of the connections means that complex clusters of atoms can be modeled by simply increasing the number of iterations, at a limited computational cost.
Finally, since the approach is only dependent on the relative positions of the atoms which determine the connections inside the cut-off radius, it is also invariant with respect to isometric transformations, i.e. reflections, translations, rotations, and combinations of those, and to permutation of atoms. 

Here, we use GNNs to explore their ability to reproduce with quantum-accuracy the potential energy surface (PES) of metals, by taking as a reference the challenging and technologically crucial example of ferromagnetic body-centered-cubic (BCC) iron. We consider {\scn} as a prototypical GNN framework that is based on the distance of atomic pairs, and we consider {\dmn} to assess the performance of a GNN scheme that also includes angular (three-body) interactions. To this purpose, we have implemented periodic boundary conditions (PBCs) and made the new {\dmn} implementation that includes PBCs available at \url{https://github.com/AilabUdineGit/GNN_atomistics/}. 
In order to machine-learn the GNN interatomic potential, we use an existing database \cite{dragoni_db} that was previously trained to develop a Gaussian Approximation Potential (GAP) \cite{PhysRevMaterials.2.013808}.

The remainder of this paper is organized as follows. 
Section Results is divided in two main subsections: 
``Implementation and training'' reports a summary of computational details, together with some performance metrics;
``Testing the GNN interatomic potentials for bcc Fe'' shows a comparative analysis of the networks based on their predictions of the properties of iron.
A general summary of the methodology and its achievements, together with suggestions for future improvements, is provided in the ``Discussion and Conclusions section''.
Finally, section Methods details the approach and the implementation of the networks, and is organized in three subsections:
``Graph Neural Networks and Message Passing'' contains a formal description of the Message Passing paradigm applied to GNNs for atomistic systems;
``Network models'' provides details of both the networks {\scn} and {\dmn}; 
``Dataset'' reports a summary of the used data.



\section*{Results}

\newcommand{\x}{\textbf{x}}
\newcommand{\e}{\textbf{e}}
\newcommand{\m}{\textbf{m}}
\newcommand{\tb}{\textbf{t}}
\newcommand{\rb}{\textbf{r}}

\subsection*{Implementation and training}



To model bulk crystal structures, the simulated atomic cluster must be embedded in an effectively infinite medium. This is achieved by using periodic boundary conditions (PBCs), which are already implemented in {\scn}. Here, PBCs have been implemented also for {\dmn}.
The training strategy is the same for both \scn\ and \dmn.
All data used for the training are from a large, existing, highly-converged DFT database \cite{dragoni_db} of bcc ferromagnetic iron that includes both pristine configurations and configurations with defects such as free surfaces, vacancies and interstitials (see Database section for details). 
A GAP potential that reproduces accurately DFT vibrational and thermodynamic properties\cite{PhysRevMaterials.2.013808} is also trained, and employed as a baseline in the comparison of the GNN models.

The training dataset is built as a subset of 80\% of the database; 
samples are randomly shuffled to avoid bias. The remaining 20\% of the samples is used to test the trained model; samples are not shuffled in this case. To regularize the distribution of the data and improve training efficiency, the per-atom energies of the whole dataset have been standardized by subtracting the mean value and dividing by the standard deviation. 
Data samples are then batched with batch size $N=6$.
A random seed is set to enable reproducibility of the process. 
The objective function, or loss, to minimize is the mean absolute error (MAE) of the difference between the predicted energy $\hat{E}_i$ and its target value $E_i$, averaged over the batch: 

  \begin{equation}
    \mathcal{L}_{MAE} = \frac{1}{N} \sum_{i=1}^N |\hat{E}_i - E_i| \ .
    \label{eq:mae}
  \end{equation}
  

For each batch, the gradient of the loss is evaluated with respect to all the trainable parameters (weights and biases) of the network. Then, the optimization algorithm minimizes the loss by adapting the parameter values. At the end of each epoch (when all the batches are evaluated) the training convergence is assessed by evaluating the MAE over all the test data.
In our setting an Adam \cite{article, loshchilov2018decoupled} optimizer was adopted. The initial learning rates, $\alpha = 10^{-4}$ for {\dmn} and $\alpha = 10^{-3}$ for {\scn}, have been fixed by performing preliminary tests.
A linear scheduler was used to reduce the learning rate if the loss did not decrease significantly; more precisely, for {\dmn} ({\scn} respectively) the learning factor is reduced by a rate of $1/10$ (respectively, $1/2$) each time the test loss was detected not to have improved by at least $1\%$ (respectively, $5\%$) over the last 10 (respectively, 3) epochs.
The more strict requirements adopted for {\scn} are due to its observed higher computational cost and difficulty for the loss to converge to the minimum.
The training is stopped when $100$ training epochs have been performed.

Using a Tesla P100 GPU with 16GB RAM, the training time amounts to $\sim{11}$ min/epoch for {\dmn} and $\sim{22}$ min/epoch for {\scn}, which means a total training time of $\sim{18}$ and $\sim{37}$ hours, respectively.
For a rough comparison, we also trained GAP on the same dataset, by using Intel Xeon E7 4860v2 CPU with $\sim{317}$GB RAM, and the training lasted $\sim{60}$ hours.
Final values of the test MAE are in the order of magnitude of tens of meV.
Inference latencies have been evaluated for 54 and 128 atoms lattices and are of the order of tens of milliseconds, with the exception of a value of 104 milliseconds for {\scn} on the smaller lattice: being a lighter model, {\scn} relies less on GPU than {\dmn} and uses only $\sim{1}\%$ of resources during 54 atoms inference, while {\dmn} uses $\sim{25}\%$.
With the more demanding 128 atoms lattice, latencies are closer and in the order of tens of milliseconds, as both the models use better the resources.
Metrics about training time, test MAE and inference latency are summarized in Table \ref{tab:efficiency}. 

\begin{table}[H]
\centering
\begin{tabular}{|c|c|c|c|} 
\hline
Metric & Unit & SchNet & DimeNet\\
\hline
Training time & min./epoch &  $\sim{22}$ & $\sim{11}$\\
Test MAE & meV & 54.8 & 23.3\\
Inference latency (54 atoms cube) & sec. & 0.104 & 0.040\\
Inference latency (128 atoms cube) & sec. & 0.041 & 0.053\\
\hline
\end{tabular}
\caption{Training time, test MAE and inference latency for {\scn} and {\dmn}.
}
\label{tab:efficiency}
\end{table}


In the original papers\cite{doi:10.1063/1.5019779, Klicpera2020Directional} atomic embeddings have size of $F=64$ for {\scn} and $F=128$ for {\dmn}.
We tested both values on both models, and obtained that while {\dmn} improves slightly from 64 to 128 (test MAE from 24.85 to 23.3), {\scn} makes a sensible leap forward (test MAE from 76.0 to 54.8).  
Consequently, an embedding size of 128 was set for both models. We consider this aspect interesting and being worth of future investigation.

The cutoff value is determined as a trade-off between two competing requirements: on one hand, the higher is the value of the cutoff, the higher is the number of connected atoms within an interaction block; on the other hand, the higher is this number, the higher is the computational cost during training. For this reason, and considering that {\dmn} is a much more complex network in which also triplets of atoms are considered, the cutoff radius is different for the two models: $r_{cut}=5.0$ {\AA} for {\scn}; $3.5$ {\AA} for {\dmn}. Using a larger cutoff (up to $4$ {\AA}) for {\dmn} did not increase the accuracy but did increase the computational time.

The presence of seven interaction blocks in {\dmn} with respect to three in {\scn} alleviates for the shorter $r_{cut}$, allowing the network to receive messages from distant atoms and to adequately model long-range interactions. 

\subsection*{Testing the GNN interatomic potentials for bcc Fe}

\begin{figure}[H]
    \centering
    \includegraphics[scale=0.45]{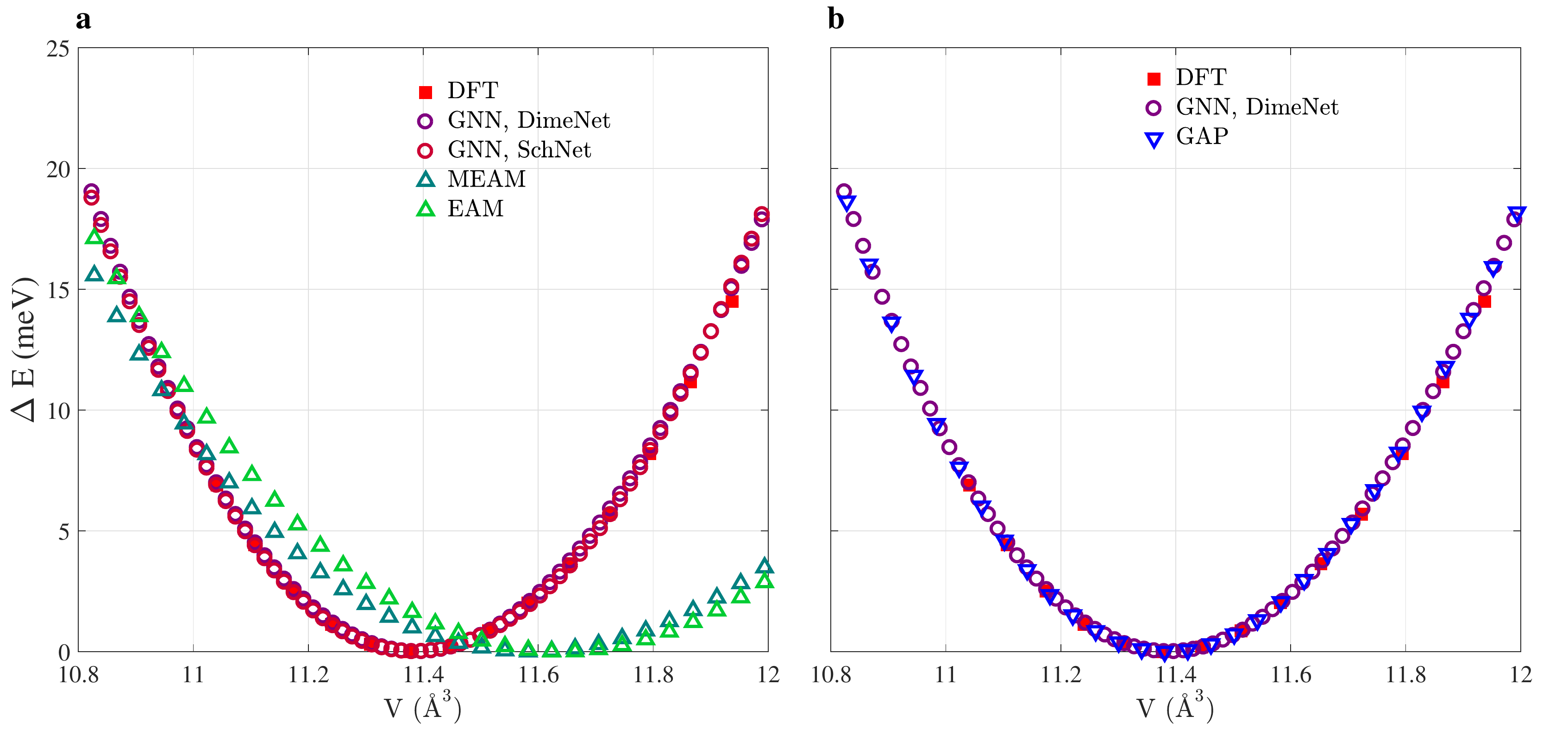}
    \caption{\textbf{a} Equation of state of SchNet and DimeNet 
    compared with DFT data and the EAM\cite{EAM_PhysRevB.79.174101} and MEAM\cite{MEAM_PhysRevB.89.094102}  empirical potentials. \textbf{b} Equation of state of the DimeNet 
    compared with DFT data and the state-of-the-art GAP iron potential \cite{PhysRevMaterials.2.013808}.}
    \label{fig:eos}
\end{figure}

The {\scn} and {\dmn} Fe potentials are benchmarked against either published DFT data\cite{PhysRevMaterials.2.013808} or data computed with Quantum Espresso based on settings (k-mesh and energy convergence) consistent with the training database \cite{PhysRevMaterials.2.013808}. 
The equation of state is computed with GNNs by varying the lattice constant $a_0 = 2.834$ {\AA} of the primitive unit cell within a range of $\pm 5\%$ volumetric change around the equilibrium volume computed with DFT. 
As shown in Fig. \ref{fig:eos}, both GNNs reproduce the DFT data with high accuracy. 

To compare the performance of GNNs with empirical potentials, we compute the equation of state, equilibrium volume and buk modulus with two broadly used empirical potentials: EAM \cite{EAM_PhysRevB.79.174101}, which is based on pairwise interactions; and MEAM \cite{MEAM_PhysRevB.89.094102}, that includes higher-order interactions (e.g. angular-dependent terms). Both GNN potentials reproduce the DFT results with high accuracy, while both the equilibrium volume and the curvature of the empirical potentials are far from the DFT results (see Fig. \ref{fig:eos}a). One reason for the discrepancy is that the empirical potentials are fitted to the experimental data of the equilibrium volume $V_0 = 11.7$ \AA$^3$, which is obtained by extrapolation to T=0K \cite{MEAM_PhysRevB.89.094102}. However, despite being fitted to such value, both EAM and MEAM visibly underpredict the experimental equilibrium volume. In contrast, both GNNs can reproduce closely the dataset they have been trained to and, as shown in Fig. \ref{fig:eos}b, the level of accuracy is comparable with the state-of-the-art GAP interatomic potential for BCC iron \cite{PhysRevMaterials.2.013808}.

The equilibrium volume and bulk modulus of iron are computed by fitting the Birch-Murnaghan equation of state to the energy-volume curve. The result of the fitting for the GNNs and DFT data is reported in Table \ref{tab:bulk}.

\begin{table}[H]
\centering
\begin{tabular}{|c|c|c|c|c|c|c|c|c|c|c|}    
\hline
Property & Unit & DFT & SchNet & $\varepsilon_{SN}$ & DimeNet & $\varepsilon_{DN}$ & GAP \cite{PhysRevMaterials.2.013808} & $\varepsilon_{GAP}$\\  
\hline

$\text{a}_0$ & \AA & $2.834$ & $2.834$ & 0.0\% & $2.834$ & 0.0\% & $2.834$ & 0.0\% \\
$\text{B}_0$ & GPa & $199.8 \pm 0.1$  & $199.0$ & -0.4\% & $199.4$ & -0.2\% & 198.2 & -0.8\% \\

\hline
\end{tabular}
\caption{T=0K lattice parameter $a_0$ and bulk modulus $B_0$ for $\alpha$-iron.
GNN results are compared to DFT data. The relative errors of \scn\ ($\varepsilon_{SN}$), \dmn\ ($\varepsilon_{DN}$) and GAP ($\varepsilon_{GAP}$) with respect to DFT are also shown.}
\label{tab:bulk}
\end{table}

As indicated by the relative errors $\varepsilon_{SN}$ and $\varepsilon_{DN}$, both \scn\ and \dmn\ reproduce the equilibrium lattice parameter and the bulk modulus with an accuracy comparable to GAP. Both the models achieve DFT-accurate results in the equation of state, with a maximum energy difference $<0.1$ meV in the volume range [11.0, 12.0] {\AA}$^3$. These results thus reveal no apparent difference between the performance of \scn\ and \dmn.

\begin{figure}[H]
    \centering
    \includegraphics[scale=0.42]{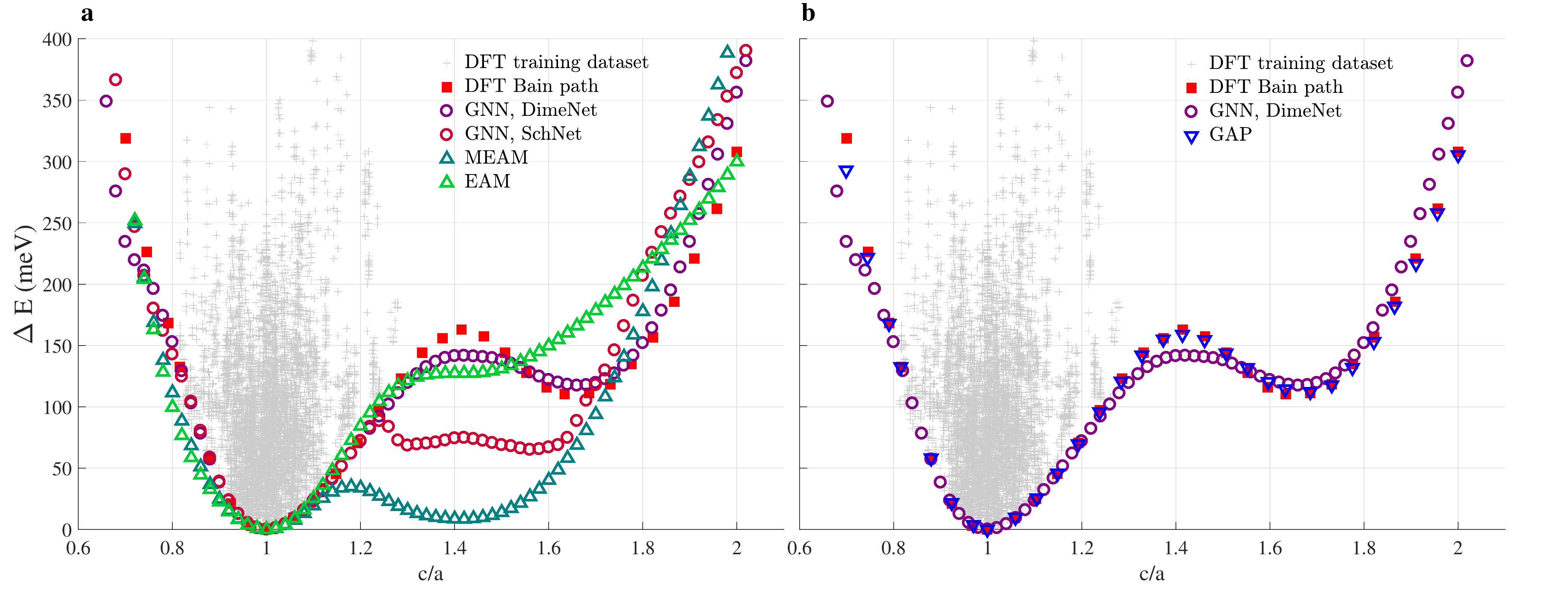}
    \caption{
    \textbf{a} DFT Bain path compared to the Bain path obtained using the GNN potentials and the empirical potentials. \textbf{b} DFT Bain path compared to the Bain path computed with DimeNet and GAP. In both panels, grey dots represent the cloud of the training data. 
    }%
    \label{fig:bain}
\end{figure}

In order to assess the ability of GNNs to reproduce tetragonal lattice distortions, the Bain path is evaluated and compared with DFT data (Figure \ref{fig:bain}).


In the figure, DFT is used to compute the energy as a function of a distorted primitive cell. The cell is distorted at constant volume, that is by increasing one axis, $c$, while reducing the two other axes, $a$, and keeping the volume constant. Both the Fe \scn\ and \dmn\ potentials are then used to compute the same path. Volume optimization, i.e. finding the minimum energy configuration at the prescribed $c/a$ by adjusting the volume, has also been performed with the GNNs potential to verify that the path does not deviate strongly from the assumed tetragonal distortion at constant volume, and no strong qualitative changes were found with respect to the result obtained with the constrained Bain path. The plot also shows the $c/a$ distortion of the training database. Fig. \ref{fig:bain}a shows that \scn\ interpolates well within the training set while it extrapolates poorly, with a discontinuous behaviour of the energy $vs$ the $c/a$ ratio. Instead, \dmn\ can extrapolate fairly well outside of the training database, also reproducing qualitatively the energy barrier at $c/a\sim1.4$ as well as the subsequent local energy minimum around $c/a\sim1.65$. This specific capability of \dmn\ sets it aside from \scn\ , making it a more promising GNN for atomistic simulations of metals with structural transformations. Moreover, \dmn\  outperforms the EAM potential, which shows no metastable minimum for BCC ferromagnetic Fe at $c/a\sim1.65$. MEAM was fitted on data including both BCC and FCC configurations, and for this reason it deviates strongly from the DFT results, which are based on BCC ferromagnetic configurations only. Fig. \ref{fig:bain}b shows that \dmn\ approaches the transferability of GAP for the Bain path.

Finally, the vacancy formation energy and the surface energies have been predicted for a number of crystal planes. The vacancy formation energy is calculated by using a $3\times3\times3$ cubic supercell. First, one atom of the supercell is removed and a DFT calculation is performed to relax the atoms around the vacancy. Then, the DFT total energy $E_{\rm def}$ of the vacancy-containing configuration is computed. The total energy $E_{\rm bulk}$ of the bulk defect-free supercell is also computed. The vacancy formation energy equals 
\begin{equation}
    E_{\rm v} = E_{\rm def} - \frac{N-1}{N} E_{\rm bulk}\ ,
\end{equation}
where $N$ is the number of atoms in the bulk system ($N=54$ atoms in this case).

\begin{figure}[H]
    \centering
    \includegraphics[scale=0.55]{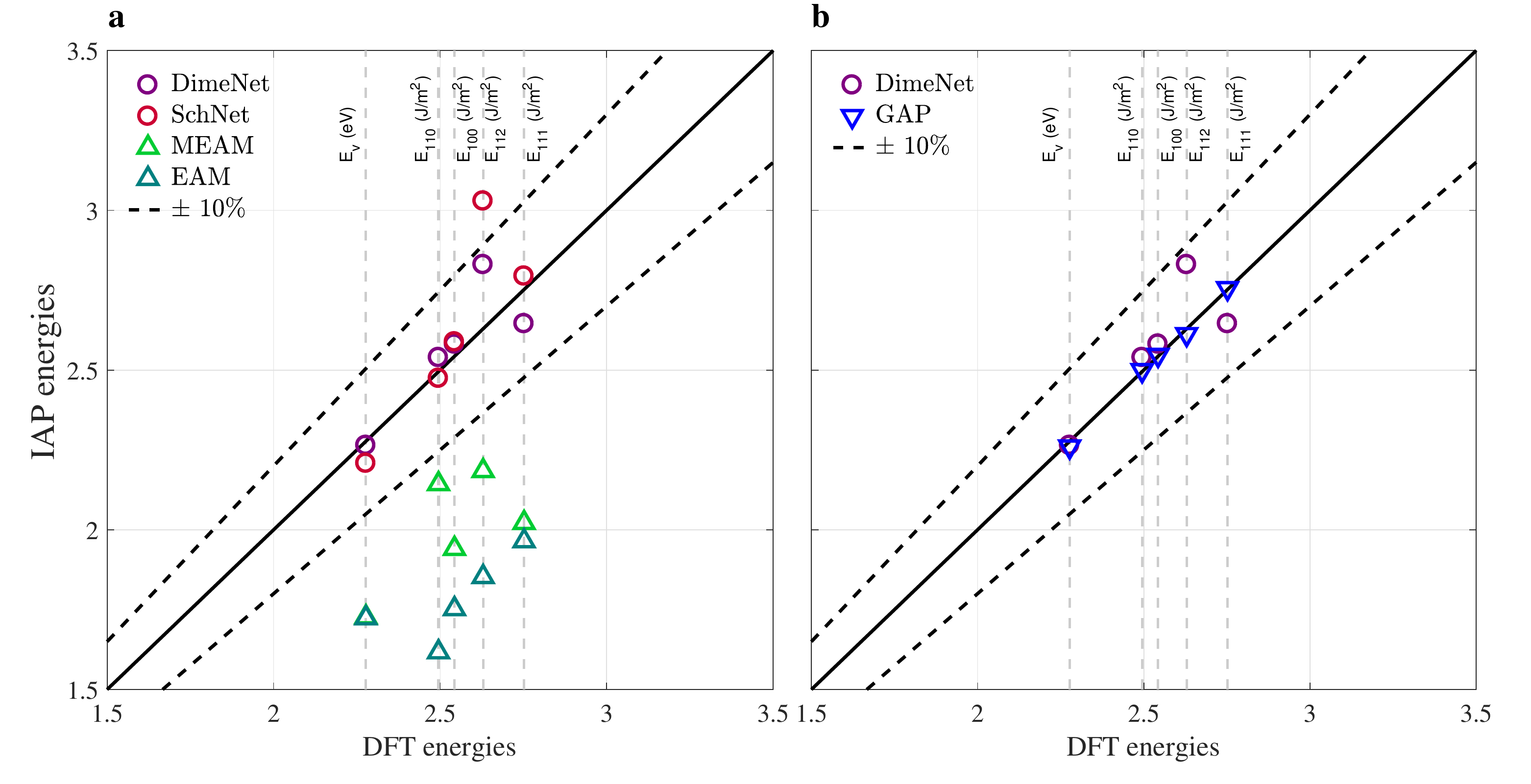}
    \caption{
    Vacancy formation energies and surface energies. The $\{100\}$, $\{110\}$, $\{111\}$ and $\{112\}$ surfaces are considered.
    }%
    \label{fig:vacsurf}
\end{figure}

The surface energy is evaluated for four crystallographic planes, i.e. $\{100\}$, $\{110\}$, $\{111\}$ and $\{112\}$. The surface is generated by creating a supercell with a vacuum region, the energy of which is indicated as $E_{\rm split}$. The vacancy formation energy is computed as

\begin{equation}
    E_{\rm surf} = (E_{\rm split} - E_{\rm bulk})/2A
\end{equation}

where $A$ is the newly created surface area. The results are plotted in Fig. \ref{fig:vacsurf}.

%
%
%


Fig. \ref{fig:vacsurf}a shows that both GNNs can reproduce mostly within 10\% accuracy all the considered DFT energies. The GNN potentials largely outperform both EAM and MEAM empirical potentials, that consistently underpredict the energies. {\dmn} shows the largest deviation ($\sim 7\%$) for the $\{112\}$ surface energy, and is otherwise approaching the predictive capabilities of the GAP potential (Fig. \ref{fig:vacsurf}b).

\section*{Discussion and Conclusions}

In this paper we presented a comparative study on the application of two GNN models, {\scn} and {\dmn}, to the prediction of properties of bcc iron.
Since {\dmn} was previously tested only on molecules and not on periodic structures/crystals, we implemented a version with PBCs and made it publicly available. Both models predict with DFT accuracy the energy-volume curve and related properties such as the bulk modulus and the equilibrium lattice parameter. This result is consistent with the fact that the energy-volume curve includes datapoints close to those of the training database. The investigated GNN potentials outperform closed-form empirical interatomic potentials (e.g. EAM and MEAM) and approach the accuracy of state-of-the-art interatomic potentials such as GAP. This makes the present GNNs implementation interesting for application to other metallic systems. 

A different performance of {\dmn} with respect to {\scn} is found for configurations including tetragonal distortions (Bain path), point defects and planar defects. {\dmn} can predict the energy of these configurations within the MAE, while the predictive capability of {\scn} is limited. We attribute this difference to the fact that, in {\dmn}, the energy depends explicitly on the angular, three-body contributions that are essential for structural transformations and for local shape distortions, while {\scn} only depends on pairwise contributions. It is also remarkable that {\dmn} has better transferability, e.g. considering the Bain path. 


By showing the capabilities of GNNs and especially the importance of three-body terms, this work supports the further investigation of GNNs and specifically {\dmn}.
Activity is currently ongoing in the following directions:
\begin{itemize}
    \item There is a number of potential improvements in terms of efficiency and accuracy of the model, which is related to the hyperparameter optimization. Further investigations will involve finding a tradeoff between chosing larger cutoff radii and/or increasing the number of interaction layers, in order to ensure the efficient description of short- and long-range interactions with high accuracy.
    \item Another aspect to be investigated is the number of features of both atom and edge embeddings, and their initialization. These are crucial characteristics in modeling the atomic environment, encoding properties such as the nature of the atom and of the pair interactions, and are expected to impact the model efficiency, e.g. in the convergence of the training.
    \item The implementation of the developed GNN Fe {\dmn} potential within the LAMMPS\cite{plimpton1995fast} open-source package is currently ongoing and will enable the systematic simulation of thermoelastic properties, as well as linear and planar defects such as dislocations and cracks that are relevant for the investigation of the mechanical properties of metals.
    \item We expect that, in the spirit of Atomic Cluster Expansions (ACE) \cite{PhysRevB.99.014104}, the transferability of GNN potentials will be improved by including more terms in the angular descriptions, by using a different choice of the radial function (e.g. based on Chebyshev polynomials), or by setting different values of the parameters $l, m$ in the angular functions (which, in the current {\dmn} implementation, are spherical harmonics with $m=0$), and/or by introducing higher-body terms, beyond the three-body term currently used in {\dmn} (see Methods). This is also the subject of current research.
\end{itemize}


\section*{Methods}



\subsection*{Graph Neural Networks and Message Passing}


A graph \cite{kipf2017semi} is a pair $\mathcal{G}=(\mathcal{V}, \mathcal{E})$ where $i \in \mathcal{V}$ are the $N$ nodes and $(i, j) \in \mathcal{E}$ are the edges. The connections among the nodes of a graph can be stored in an adjacency matrix $A \in \mathbb{R}^{N \times N}$ containing the pairs $(i, j)\in \mathcal{E}$. At both nodes and edges, vectors of features (or embeddings) are defined as $\x_i \in \mathbb{R}^F$ and $\e_{ij} \in \mathbb{R}^D$, respectively, where $F, D$ are model specific parameters. 
In the message passing with node update, node embeddings are updated 
iteratively, with each iteration executed in the message passing layers $l$ as follows:


  \begin{equation}
    \x^{(l+1)}_i = \gamma (\x^{(l)}_i, \sum_{j\in \mathcal{N}_i} \mu (\e^{(l)}_{ij}, \x^{(l)}_j))
    \label{eq:node-update}
  \end{equation}

where $\mathcal{N}_i$ is the set of the nodes connected to node $i$, $\mu$ is a differentiable function of the nodal and edge embeddings, the sum aggregates the contributions of atoms $j$,
and $\gamma$ is a differentiable function which evaluates the update of node embedding. 

In the message passing with edge update \cite{31bef22ac7034baca72d1f08d3b16c4b}, edge embeddings are updated by following a similar scheme:


  \begin{equation}
    \e^{(l+1)}_{ij} = \kappa (\e^{(l)}_{ij}, \sum_{k\in \mathcal{N}_j \setminus\{i\}} \nu (\x^{(l)}_j, \e^{(l)}_{jk}, \x^{(l)}_k))\ .
    \label{eq:edge-update}
  \end{equation}
  
with the same conventions as the previous case, $\kappa$ and $\nu$ being differentiable functions of the nodal and edge embeddings, analogously to $\gamma$ and $\mu$. Note that edges connected to $(i, j)$ are the edges $(j, k)$ linking node $j$ and node $k \neq i$, hence the index of the summation. 

\begin{figure}[H]
  \centering
  \includegraphics[width=.4\textwidth]{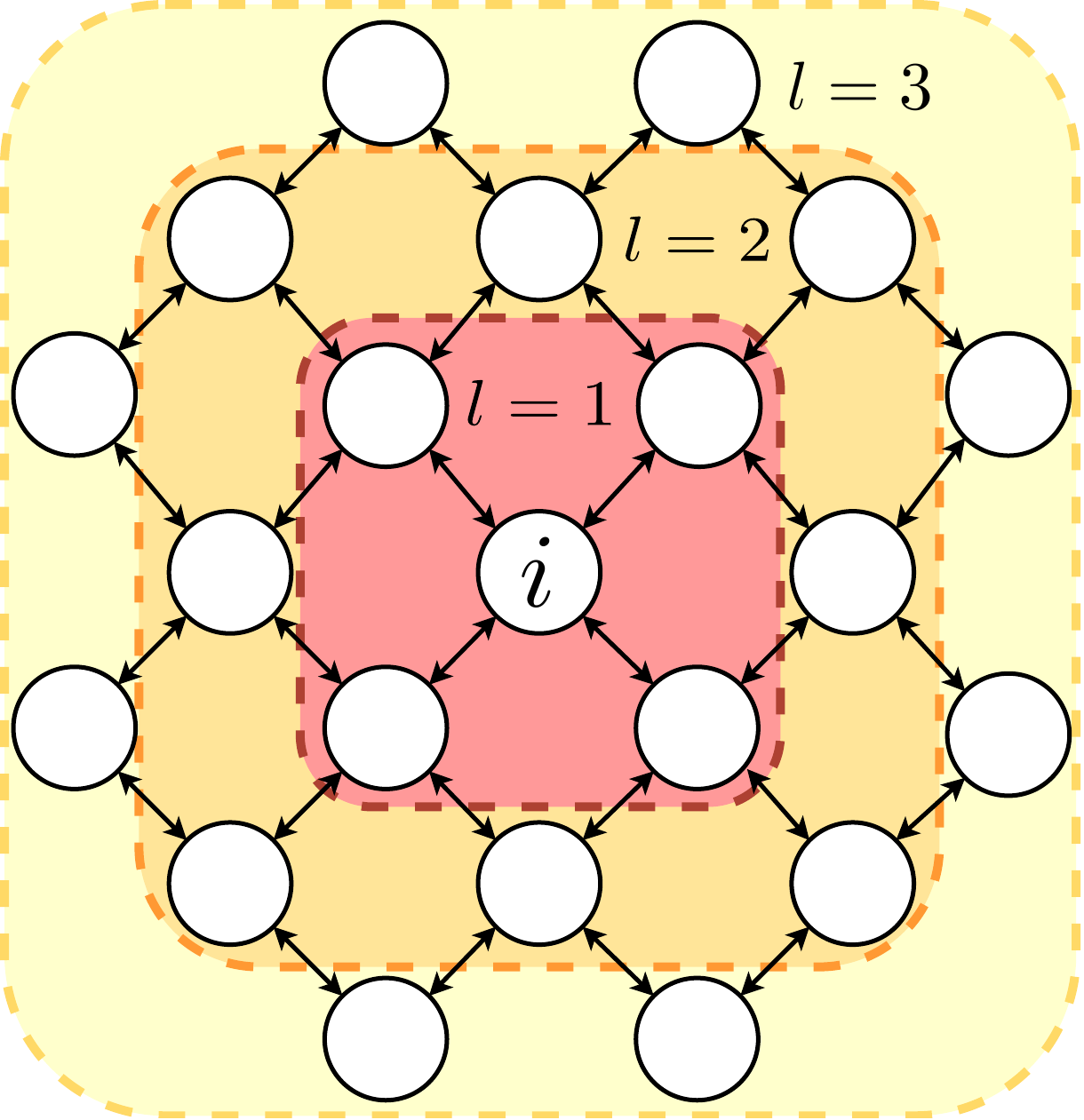}
  \caption{
  Message Passing with node update. Atomic environment as seen by node $i$:
  layer $l=1$ aggregates messages from connected neighbours (red area); 
  layer $l=2$ acts the same but connected neighbours have been already updated by messages from their neighbours at previous layer, so signals received by atom $i$ are now from a distance of up to two edges (orange area). A sequence of $L$ layers means messages coming from nodes at a distance of up to $L$ edges.
  A similar scheme works for message passing with edge update.
  }
  \label{fig:message_passing}
\end{figure}

At the next iteration, the message is evaluated by the layer $l+1$ by aggregating embeddings $\x^{(l+1)}_i$ ($\e^{(l+1)}_{ij}$) from the neighbours, which in turn have received a message from their own neighbours: stacking together $L$ layers means that the final update is performed by using messages coming from a distance up to $L$ neighbors away, see fig. \ref{fig:message_passing}. 

Once iteratively updated via the message passing, embeddings are elaborated by a readout function 

  \begin{equation}
    y = f (\{\x^{(L)}_i, \e^{(L)}_{ij}\})\
    \label{eq:readout}
  \end{equation}
  
which performs a further aggregation of all the embeddings and outputs the prediction $y \in \mathbb{R}$ of the network.

Within the context of crystalline materials, it is straightforward to consider the nodes of a graph as the atoms and to connect by edges the pairs of atoms that lie within a specified interaction radius. Let $\rb_i \in \mathbb{R}^3$ be the coordinates of the atom $i$. Then, the graph is defined by connecting all the atoms $j$ that are inside the cutoff radius $r_{cut}>||\rb_i - \rb_j||$.
Atomic embeddings $\x_i$ are vectors of learnable numerical features.
They are randomly initialized, and atoms with the same set of relevant atomic properties, such as atomic number $Z$, have the same initial embeddings.
Edge embeddings $\e_{ij}$ are similar, with properties related to pairs of connected atoms, such as the interatomic distance $d_{ij}$. 
The message and update functions $\mu, \gamma$, (\ref{eq:node-update} \ref{eq:edge-update}) are neural layers which add learnable weights and define the form of the convolutional filter and of the update procedure.
Hence, the prediction of the potential energy $E \in \mathbb{R}$ of a crystal as a function of the atomic coordinates is a regression task performed on such a graph (\ref{eq:readout}). 

\subsection*{Network models}
\label{subsection:network_models}

\begin{figure}[H]
  \centering
  \includegraphics[width=.9\textwidth]{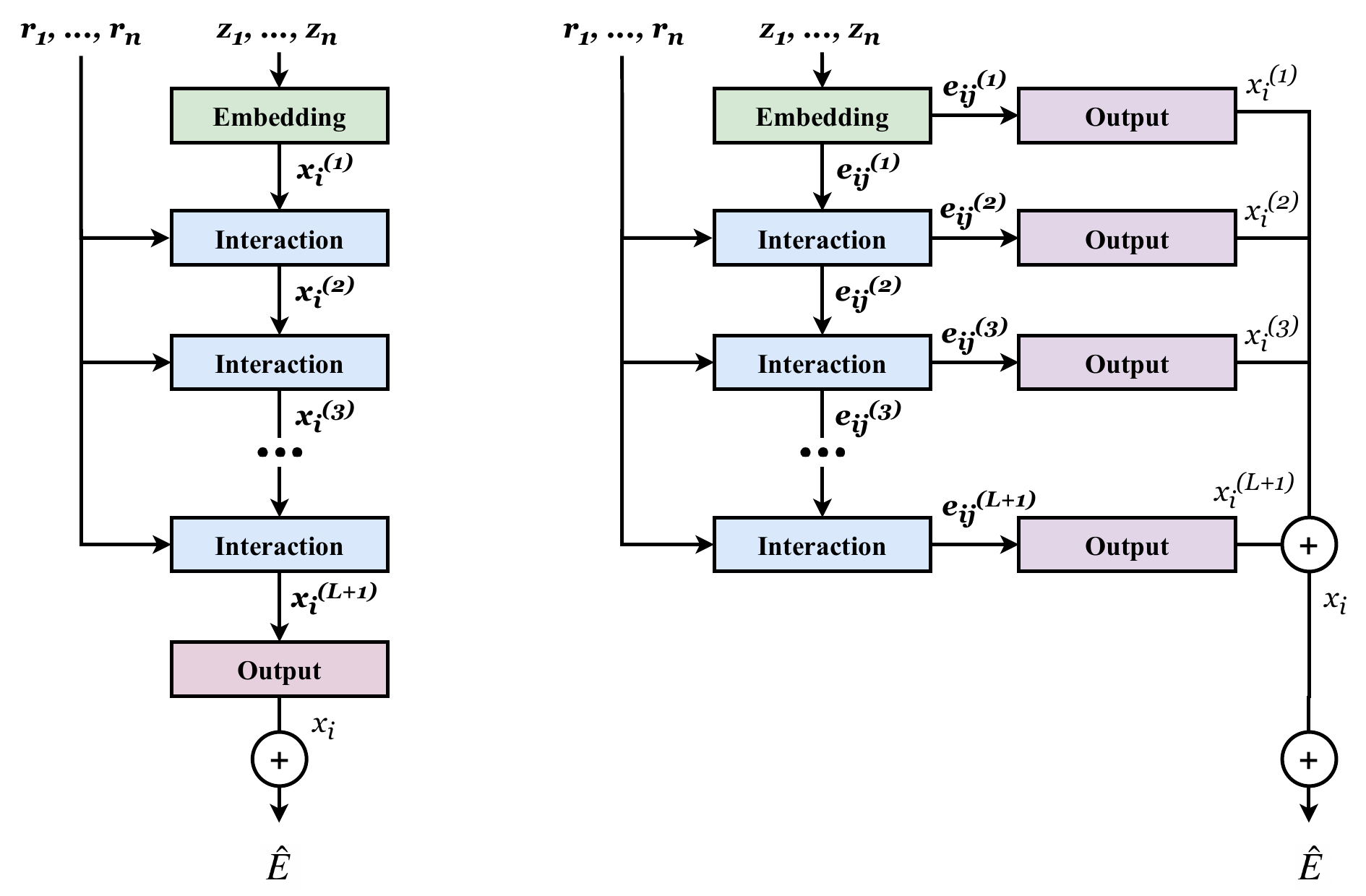}
  \caption{
  Block diagram for {\scn} (left) and {\dmn} (right).
  Outputs generated by each block are shown near the arrows.
  In both models the starting point is the Embedding block that maps atom or edge features in a vector space, generating embeddings.
  For {\scn}, the Interaction blocks are in a sequence, the output of one being the input of the next; the final Output block evaluates the total energy $\hat{E}$.
  In {\dmn} the output of each block is both passed sequentially to the next and further elaborated by the Output block, and then summed to obtain energy $E$. 
  }
  \label{fig:models_schema}
\end{figure}

In this paper we use two recent models of Graph Neural Networks based on the Message Passing framework: \scn \cite{doi:10.1063/1.5019779} and \dmn \cite{Klicpera2020Directional}.
There are two main differences between them. The first is related to the embeddings: {\scn} relies on atomic embeddings, while {\dmn} also uses edge embeddings in the form of pairs of atom embeddings to account for the directionality of the message passing. The second difference is the learned convolutional filters used to aggregate embeddings: while {\scn} employs a filter that accounts only for the distance between pairs of atoms, {\dmn} considers also the angles formed by pairs of edges, or triplets of atoms.
The general scheme of both the models is shown in fig. \ref{fig:models_schema}. At an high level of abstraction, they can be described in terms of block diagrams, with each block representing a set of specific neural layers that performs some operations on input data and generates output data.
Blocks with the same name in both models perform similar general operations.


\subsubsection*{Filters and physical representation of the atomic environment}
\label{subsubsection:distances-angles}

To take into account the physical environment surrounding each atom, the convolutional filter assigns weights to the embeddings received by the neighbours (see below the description of Embedding and Interaction blocks). 
Filter weights are learned during training and have to change smoothly with respect to small atomic shifts.
Therefore, distances and angles are {\it expanded}, that is represented as feature vectors whose components are sets of continuous basis functions.
In {\scn} the filter depends only on the interatomic distance $d$, 
expanded by a set of radial, Gaussian (G) basis functions:

  \begin{equation}
    \phi^G_k(d) = \exp \left(-\frac{\left(d - \mu_k\right)^2}{2 \sigma^2}\right)
    \label{eq:sch-filter}
  \end{equation}

with $\mu_k$ equally spaced in the interval $[0, r_{cut}]$, and $\sigma$ representing the scale of the distances. 
Hyperparameters $k$ and $\sigma$ define the granularity of the representation, and determine the precision of the filter. 
The spacing $r_{cut}/k$ and the scale $\sigma$ are both set to 0.1 {\AA} in the original paper \cite{doi:10.1063/1.5019779}; in order to improve the precision to better compare with {\dmn} we set them to 0.04 {\AA}.

{\dmn} introduces two different filters: one radial depending only on distances, used to weight the embeddings received by atoms; 
and one radial-angular which takes into account also the angles to weight the embeddings passed to the edges. 
Both distances and angles are expanded in a 2D Bessel-Fourier basis which are the solutions of the related time-independent Schr\"odinger equation, and represent the electron density of the system inside the cutoff radius.
\ For the first, only radial, filter, distances $d$ are expanded in a feature vector whose components are given by the Radial Basis Functions (RBF):

  \begin{equation}
    \phi^{RBF}_k(d) = \sqrt{\frac{2}{r_{cut}}} \frac{\sin{\left( \frac{k\pi}{r_{cut}} d\right)}}{d} \ .
    \label{eq:dist-filter}
  \end{equation}

The second filter depends on the distance $d$ and the angle $\theta$. The components of the bidimensional feature vectors are given in terms of the Spherical Basis Functions (SBF):

  \begin{equation}
    \phi^{SBF}_{l,k}(d, \theta) = \sqrt{\frac{2}{r_{cut}^3 j^2_{l+1}\left(z_{lk}\right) }} j_l \left( \frac{z_{lk}}{r_{cut}} d\right) Y^0_l (\theta)
    \label{eq:angle-filter}
  \end{equation}

where $j_l$ are the spherical Bessel functions of the first kind and $Y^m_l$ are the spherical harmonics; $z_{lk}$ is the $k$-th root of the $l$-order Bessel function. 
Settings for the non learnable parameters are as per the original paper \cite{Klicpera2020Directional}, namely: for eq. \ref{eq:dist-filter} $k \in [1, \dots, 6]$ while in eq. \ref{eq:angle-filter} $k \in [1, \dots, 6]$, $l \in [0, \dots, 5]$.
To avoid the discontinuity given by the boundary condition $\phi(d) = 0$ for $d > r_{cut}$, functions \ref{eq:dist-filter} and \ref{eq:angle-filter} are multiplied by a smoothing polynomial $u(d) \sim \mathcal{O}(d^8)$: a step function with a root of multiplicity 3 at $d = r_{cut}$. 

\begin{figure}[H]
  \centering
  \includegraphics[width=\textwidth]{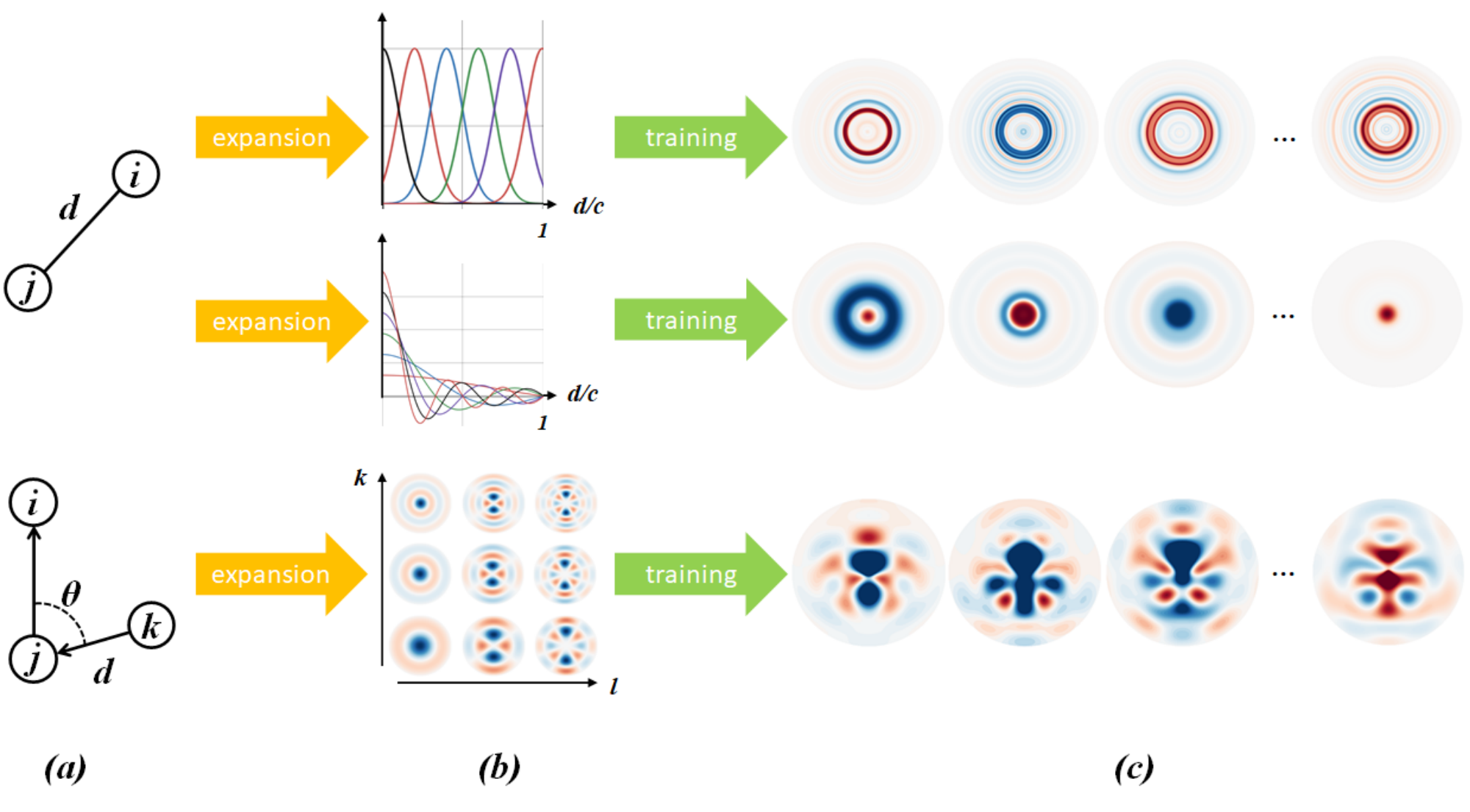}
  \caption{
  Filters and physical representation of the atomic environment.
  (a) Starting from the positions of atoms, distances between pairs (top, SchNet and DimeNet) and distances and angles between triplets (bottom, DimeNet) are evaluated.
  (b) Distances $d$ are expanded in a Gaussain basis of functions (top, SchNet, eq. \ref{eq:sch-filter}) or in Radial Bessel basis (middle, DimeNet, eq. \ref{eq:dist-filter}), while distances $d$ and angles $\theta$ for triplets are expanded in a 2D Bessel-Fourier basis (bottom, DimeNet, \ref{eq:angle-filter}). 
  (c) The convolutional filters: expansions are passed through dense layers whose weights are optimized during training. Learned weights are the convolutional filters. The first three and the last component are extracted and shown for all the cases.
  }
  \label{fig:filters}
\end{figure}

For both {\scn} and {\dmn} the expanded representations are passed through dense neural layers (see below) which add the learnable weights.
The filter is therefore a mapping of the physical representation of angles and distances to a vector space with dimensions matching the ones of the embeddings to weight.
The general aspect of the filters and an intuition of how they work are shown in fig. \ref{fig:filters}.

\subsubsection*{Dense layers}
Dense layer is the very basic element of a neural network. Given an input $\x \in \mathbb{R}^k$ it is defined as

  \begin{equation}
    \textbf{y} = \sigma (\textbf{W} \cdot \x + \textbf{b})
    \label{eq:dense}
  \end{equation}
  
where $\textbf{W} \in \mathbb{R}^{m \times k}$, $\textbf{b} \in \mathbb{R}^m$ are the learnable weights and bias, $\cdot$ is the matrix multiplication operator and $\sigma$ is the \emph{activation}, i.e. a differentiable nonlinear function.
Activation is the shifted softplus for \scn: $ssp(x) = \ln (0.5\cdot e^x + 0.5)$, and the self-gated Swish for \dmn: $sgs(x) = x \cdot \sigmoid(x)$.
In terms of vector algebra a dense layer projects the input vector $\x \in \mathbb{R}^k$ into a vector space $\mathbb{R}^m$ with $m \neq k$ in general, and then applies the function $\sigma$ element-wise.


\subsubsection*{Embedding block}
For \scn, atom embeddings are defined as vectors $\x_i \in \mathbb{R}^F$; initial values $\x_i^{(0)}$ are randomly initialized.
For \dmn, similarly defined atom embeddings are concatenated in pairs to generate an initial edge embedding $\e_{ji}^{(0)} = (\x_j^{(0)} || \x_i^{(0)} || \phi^{RBF}_k(d_{ji}))$. Note that this definition guarantees the directionality, as in general $\e_{ji} \neq \e_{ij}$.
Once initialized, embeddings are passed through dense layers.


\subsubsection*{Interaction block}
Message passing paradigm is implemented in Interaction blocks.
Multiple Interaction blocks are generally stacked together. Each of them performs a convolution by aggregating embeddings from the directly connected entities, and then updating them.
The output of one block is passed as the input to the next.

\begin{figure}[H]
  \centering
  \includegraphics[width=.5\textwidth]{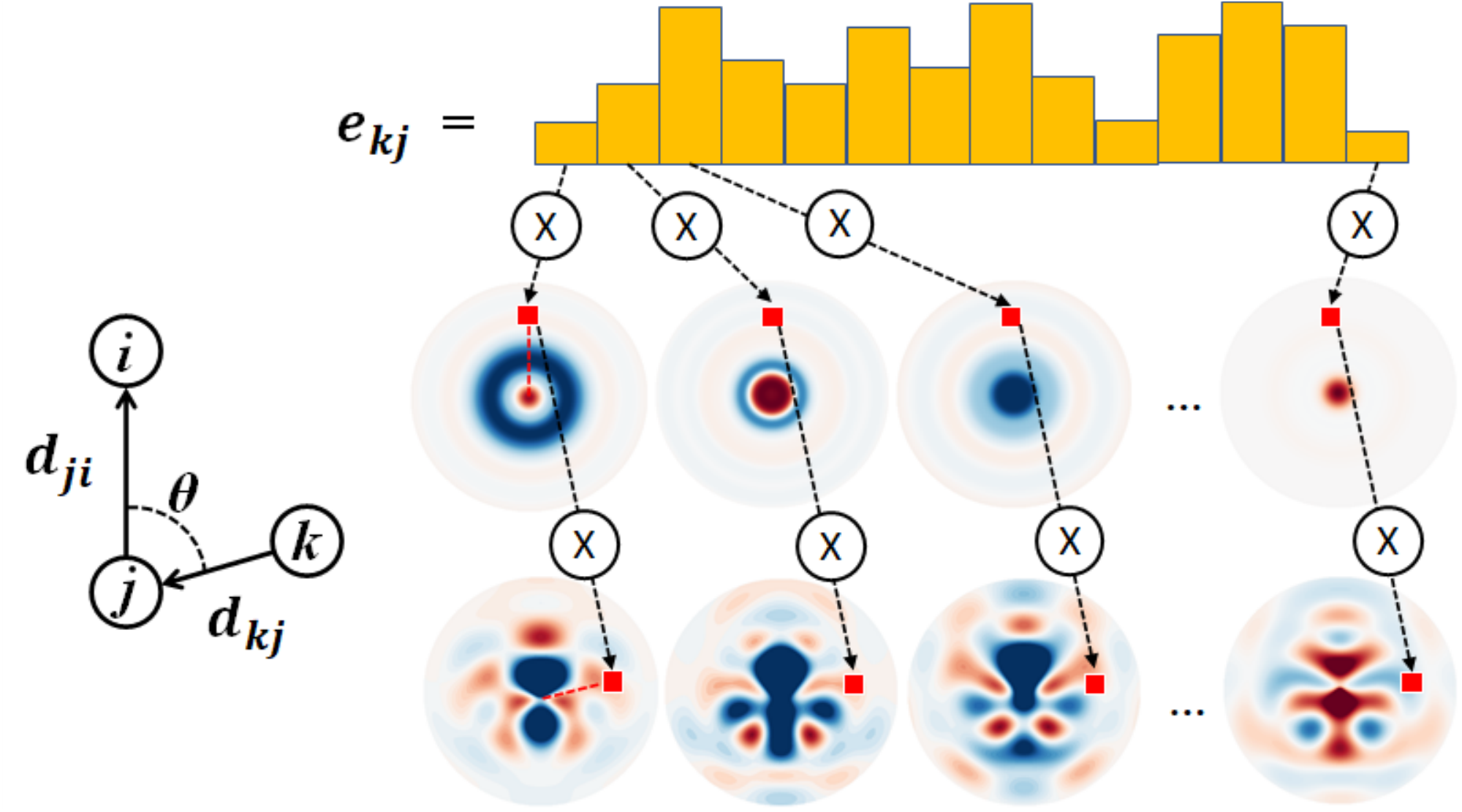}
  \caption{
  Schematic of the application of the filters for {\dmn}.
  Features of the edge embedding $e_{kj}$ are first multiplied element-wise with the value at the point $d_{ji}$ of the components of the radial filter.
  Then another element-wise multiplication is performed with the value at the point $(d_{kj}, \theta)$ of the components of the radial-angular filter.
  An analogous scheme works for {\scn}, limited to the purely radial filter.
  }
  \label{fig:filters_application}
\end{figure}

Let $l$ be the generic Interaction block. In {\scn}, the embedding $\x_j^{(l)}$ received by atom $i$ from neighbour $j \in \mathcal{N}_i$ is first weighted by the gaussian radial filter depending on $\phi^G(d)$ (eq. \ref{eq:sch-filter}). Then embeddings are aggregated and the resulting embedding is summed to $\x_i^{(l)}$ and passed through a dense layer to update it to $\x_i^{(l+1)}$ (eq. \ref{eq:node-update}).
In {\dmn}, the edge $(j, i)$ receives message embeddings $\e_{kj}^{(l)}$ from edges $(k, j)$ that are first weighted by means of the radial filter based on $\phi^{RBF}(d)$ (eq. \ref{eq:dist-filter}) with $d = d_{ji}$, and then by the radial-angular filter based on the Bessel-Fourier basis $\phi^{SBF}(d, \theta)$ (eq. \ref{eq:angle-filter}), where $\theta$ is the angle formed by $(j, i)$ and $(k, j)$ and $d = d_{kj}$.
Again, exchanged messages are aggregated, summed over $k$ to the embedding $\e_{ji}^{(l)}$ relative to edge $(j, i)$ and then passed through the dense layers to obtain the updated $\e_{ji}^{(l+1)}$ edge embedding (eq. \ref{eq:edge-update}).
For an intuition of how the filters are applied see fig. \ref{fig:filters_application}.
In {\dmn}, updated messages are given as input to the next interaction block \emph{and} to the related output block, see below.


\subsubsection*{Output block}
Output blocks are responsible for lowering the dimensions of the atom embeddings, reducing them to scalars.

\scn~ has one only Output block at the end of the Interaction blocks stack; it is a sequence of dense layers whose task is to reduce the embedding to a scalar $\x_i^{(L+1)} \rightarrow x_i^{(L+1)}$, to be interpreted as the atom-wise contribution to the total potential energy.
The final prediction is evaluated as the sum of atom-wise contributions $\hat{E} = \sum_i x_i^{(L+1)}$.

\dmn~ performs another convolution here, resulting in the update of the atomic embeddings.
Embeddings $\e_{ij}^{(l+1)}$ from the related interaction block $l$ (and of the embedding block, $l=0$) are further convoluted by means of a radial filter based on $\phi^{RBF}(d)$: $\e_{ij}^{(l+1)} \rightarrow \Tilde{\e}_{ij}^{(l+1)}$. 
The update of the embedding of atom $i$ is then evaluated as  $\x_i^{(l+1)} = \sum_j \Tilde{\e}_{ij}^{(l+1)}$. 
Multiple dense layers are applied to reduce dimensions to 1: $\x_i^{(l+1)} \rightarrow x_i^{(l+1)}$, to be intended as the per-atom contribution of the level $l$ blocks to the output of the model.
Finally they are summed atom-wise and level-wise to evaluate the final prediction of the network $\hat{E} = \sum_l \sum_i x_i^{(l+1)}$. 



\subsection*{Dataset}
\label{subsection:dataset}

We use a DFT database\cite{PhysRevMaterials.2.013808} of bcc ferromagnetic iron in our study. The database is generated by delicate collinear spin-polarized plane wave DFT computations, which includes the following subsets. 
\begin{itemize}
    \item DB1: Primitive unit cell under arbitrary pressures at 300K
    \item DB2: $3\times 3\times 3$ and $4\times 4\times 4$ supercell under a range of pressures and temperatures
    \item DB3: $3\times 3\times 3$ supercell containing a vacanvy under a range of pressures and temperatures the same as DB2
    \item DB4: $4\times 4\times 4$ supercell with divacancies at 800K
    \item DB5: $4\times 4\times 4$ supercell with 3, 4 and 5 vacancies at 800-1000K
    \item DB6: $4\times 4\times 4$ supercell containing self- and di-interstitials at 100-300K
    \item DB7: $1\times 1\times 6$ supercell with (100), (110), (111) and (112) free surfaces 
    \item DB8: $1\times 1\times 6$ supercell with $\gamma$ surfaces on (110) and (112) plane
\end{itemize}
All structures in DBs other than DB1 are bcc lattices; structures in DB1 are primitive unit cells of bcc lattices.
More details about the database can be found in the original paper \cite{PhysRevMaterials.2.013808}. The DFT database is computed by using the open source codes QUANTUMESPRESSO\cite{Giannozzi_2009,Giannozzi_2017}. An ultrasoft GGA PBE pseudopotential from $0.2.1$ pslibrary is employed. The kinetic energy cutoff for wavefunctions and charge density are set to be $90$ and $1080$ $Ry$, respectively. The $k$ spacing is set to be less than $0.03$ \AA$^{-1}$. 


\section*{Data and code availability}

The data used for training and testing the system is publicly available at the \emph{Materials Cloud} site: \url{https://archive.materialscloud.org/record/2017.0006/v2}.
The code generated to obtain the data reported in the paper can be found at the GitHub repository of the project: \url{https://github.com/AilabUdineGit/GNN_atomistics/} 

\bibliography{main}

\end{document}